\documentclass[a4paper,aps,twocolumn,nofootinbib]{revtex4}
\RequirePackage[colorlinks,hyperindex]{hyperref}
\RequirePackage[english]{babel}
\RequirePackage[latin1]{inputenc}
\RequirePackage[T1]{fontenc}
\RequirePackage{mathrsfs}
\RequirePackage{amsmath}
\RequirePackage{amssymb}
\RequirePackage{amsbsy}
\RequirePackage{color}
\RequirePackage{bm}
\hypersetup{colorlinks=true,breaklinks=true,urlcolor=blue,linkcolor=red}
\begin{document}
\title{\bf{The Tensorial Connections}}
\author{Luca Fabbri}
\affiliation{DIME, Sezione di Metodi e Modelli Matematici,\\
Universit\`{a} di Genova, Via all'Opera Pia 15, 16145 Genova, ITALY}
\date{\today}
\begin{abstract}
In a series of recent papers, we have introduced an object that was constructed on the connection but which was proven to be a tensor: this object, thus called tensorial connection, has been defined and some of its properties have been given. In the present paper, we intend to present all the results found so far, complementing them with some new ones, in a systematic and organic manner.
\end{abstract}
\maketitle
\section{Introduction}
General covariance is the principle stating that any geometry must be built while respecting some fundamental symmetries of the background. Hence, for the space-time general covariance requires the invariance under the most general coordinate transformation, and the objects that comply with this property are the tensors. Nevertheless, differentiation is not compatible with general covariance unless an additional structure called connection is introduced. This connection is defined in terms of a transformation whose non-tensorial character is specifically given so as to compensate the non-tensorial character of partial differentiation, thus resulting into an improved derivation that does respect general covariance. To interpret these facts, we may think that non-linear coordinates produce artificial distorsions in the tissue of the space-time which must be straightened before taking the incremental ratio needed to calculate the derivative, and this straightening is performed by the connection. Thus, the connection is where we find information about non-linear distorsions of the space-time, like inertial acceleration. And when such distorsions are not artifacts of non-linear coordinates but real curvatures of the space-time, then the accelerations are not due to inertial forces but to gravitational ones.

The introduction of ortho-normal bases of tetrad fields brings some novelty into this picture, because tetrads are defined by two transformations. As said above, since they are vectors, their components have to respect general covariance under coordinate transformation. But additionally, they are a basis, so they can be linearly recombined with some Lorentz transformation. Hence, the existence of bases of tetrads allows to have general covariance under coordinate transformations converted into general covariance under Lorentz transformations, or passive transformations converted into active transformations. In doing so, the connection converts into the spin connection, being a vector for passive transformations with non-tensor character for active transformations. It is within the spin connection that now lies all information about non-linear distorsions, giving inertial accelerations. And in the curvature of space-time still are accelerations due to gravity.

The specific form of the active transformations codified through the Lorentz group makes it possible to search for representations different from the real one and the other possibility is to look for representations that are complex, called spinorial. Spinorial transformations are what can define spinorial fields. Derivatives can also be defined in such a way as to respect spinorial covariance if spinorial connections are introduced. Spinorial connections can be defined in most general terms. Nevertheless, they always admit one decomposition in terms of the spin connection that is unique up to the addition of specific extra fields, as we are going to discuss next. Just the same, the spinorial connection is where we find the information about inertial acceleration. In its curvature is where we eventually find the information about accelerations due to gravitation.

This is the general background. However, very recently some development was added in view of the so-called polar form of spinorial fields. Spinor fields are almost always treated in terms of their plane-wave form, but clearly a general mathematical study is possible \cite{G, L, Cavalcanti:2014wia, HoffdaSilva:2017waf, daSilva:2012wp, daRocha:2008we, Villalobos:2015xca, Ablamowicz:2014rpa, Vignolo:2011qt, daRocha:2013qhu, Ahluwalia:2004sz, Ahluwalia:2004ab, Ahluwalia:2016rwl, Ahluwalia:2016jwz, daRocha:2007pz, HoffdaSilva:2009is, Cavalcanti:2014uta, Bernardini:2005wh, Bernardini:2012sc, daRocha:2011yr, Rodrigues:2005yz}. In such a general mathematical study, having spinors in polar form means to have their components written as a real module times a unitary complex phase while still respecting their spinorial covariance, with the advantage that the spinor components are re-arranged so to manifest what are the degrees of freedom \cite{Fabbri:2016msm, Fabbri:2016laz, Fabbri:2017pwp}. But there is more to it, since in polar form all the components that are not degrees of freedom concur together with the spin connection to form quantities that, while containing the same information of the connection itself, nevertheless are true tensors. Such a tensorial connection encodes information of a covariant type of inertial acceleration, since it is generally non-zero but nevertheless curvatureless \cite{Fabbri:2018crr}. And one can imagine it as some sort of tension of the space-time that may have effects like the localization of matter distributions \cite{Fabbri:2019kfr}.

In \cite{Fabbri:2018crr, Fabbri:2019kfr} we have argued that an inertial acceleration that can never be vanished by a choice of reference system may not necessarily be a strange concept since we already know of the existence of an angular momentum that can never be vanished by a choice of frame. This is the spin, and covariant inertial accelerations may simply be sorts of accelerations related to spin. Neither can be vanished by choice, both seem defined only for a spinor. However, in the following we will show that it is possible to define tensorial connections in more general circumstances.
\section{Fundamental Geometry}
To begin, we recall the general mathematical concepts.

The space-time metric $g_{\mu\nu}$ is diagonalized in terms of the tetrad fields $e_{a}^{\mu}$ according to $e_{a}^{\mu}e_{b}^{\nu}g_{\mu\nu}\!=\!\eta_{ab}$ where $\eta_{ab}$ is the Minkowski metric. Matrices $\boldsymbol{\gamma}_{a}$ verify the relations given by $\{\boldsymbol{\gamma}_{a} ,\!\boldsymbol{\gamma}_{b}\}\!=\!2\mathbb{I}\eta_{ab}$ (known as Clifford algebra).

In terms of all these objects, it is possible to define
\begin{eqnarray}
&(\sigma_{ab})^{i}_{\phantom{i}j}\!=\!\delta^{i}_{a}\eta_{jb}\!-\!\delta^{i}_{b}\eta_{ja}
\end{eqnarray}
as the generators of the Lorentz algebra in real representation, so that with parameters $\theta_{ab}\!=\!-\theta_{ba}$ we build
\begin{eqnarray}
&\Lambda\!=\!e^{\frac{1}{2}\theta_{ab}\sigma^{ab}}
\end{eqnarray}
as Lorentz transformation in real representation. In very similar ways we can prove that 
\begin{eqnarray}
&\boldsymbol{\sigma}_{ab}\!=\!
\frac{1}{4}\left[\boldsymbol{\gamma}_{a},\!\boldsymbol{\gamma}_{b}\right]
\end{eqnarray}
are the generators of the Lorentz algebra in complex representation, with the same $\theta_{ab}\!=\!-\theta_{ba}$ we get 
\begin{eqnarray}
&\boldsymbol{S}\!=\!e^{\frac{1}{2}\theta_{ab}\boldsymbol{\sigma}^{ab}}
\end{eqnarray}
as Lorentz transformation in complex representation, or spinorial representation, and it is easy to prove that this representation is reducible since all generators commute with the $\boldsymbol{\pi}$ matrix defined as $2i\boldsymbol{\sigma}_{ab}\!=\!\varepsilon_{abcd}\boldsymbol{\pi}\boldsymbol{\sigma}^{cd}$ (we stress that this matrix is what is usually indicated as a gamma with an index five which is meaningless in the space-time, and therefore we prefer to use a notation with no index).

Clifford matrices verify the relationships
\begin{eqnarray}
&\boldsymbol{\gamma}_{i}\boldsymbol{\gamma}_{j}\boldsymbol{\gamma}_{k}
\!=\!\boldsymbol{\gamma}_{i}\eta_{jk}-\boldsymbol{\gamma}_{j}\eta_{ik}
\!+\!\boldsymbol{\gamma}_{k}\eta_{ij}
\!+\!i\varepsilon_{ijkq}\boldsymbol{\pi}\boldsymbol{\gamma}^{q}
\end{eqnarray}
from which it is possible to prove that
\begin{eqnarray}
&\{\boldsymbol{\gamma}_{a},\boldsymbol{\sigma}_{bc}\}
=i\varepsilon_{abcd}\boldsymbol{\pi}\boldsymbol{\gamma}^{d}\\
&[\boldsymbol{\gamma}_{a},\boldsymbol{\sigma}_{bc}]
=\eta_{ab}\boldsymbol{\gamma}_{c}\!-\!\eta_{ac}\boldsymbol{\gamma}_{b}\label{commgamma}
\end{eqnarray}
and then deduce that also
\begin{eqnarray}
&\{\boldsymbol{\sigma}_{ab},\boldsymbol{\sigma}_{cd}\}
=\frac{1}{2}[(\eta_{ad}\eta_{bc}\!-\!\eta_{ac}\eta_{bd})\mathbb{I}
\!+\!i\varepsilon_{abcd}\boldsymbol{\pi}]\\
&[\boldsymbol{\sigma}_{ab},\boldsymbol{\sigma}_{cd}]
=\eta_{ad}\boldsymbol{\sigma}_{bc}\!-\!\eta_{ac}\boldsymbol{\sigma}_{bd}
\!+\!\eta_{bc}\boldsymbol{\sigma}_{ad}\!-\!\eta_{bd}\boldsymbol{\sigma}_{ac}\label{commsigma}
\end{eqnarray}
are valid as general geometric spinorial matrix identities.

Given the spinor field $\psi$ its complex conjugate spinor field $\overline{\psi}$ is defined in such a way that bi-linear quantities
\begin{eqnarray}
&\Sigma^{ab}\!=\!2\overline{\psi}\boldsymbol{\sigma}^{ab}\boldsymbol{\pi}\psi\\
&M^{ab}\!=\!2i\overline{\psi}\boldsymbol{\sigma}^{ab}\psi
\end{eqnarray}
with
\begin{eqnarray}
&S^{a}\!=\!\overline{\psi}\boldsymbol{\gamma}^{a}\boldsymbol{\pi}\psi\\
&U^{a}\!=\!\overline{\psi}\boldsymbol{\gamma}^{a}\psi
\end{eqnarray}
as well as
\begin{eqnarray}
&\Theta\!=\!i\overline{\psi}\boldsymbol{\pi}\psi\\
&\Phi\!=\!\overline{\psi}\psi
\end{eqnarray}
are all real tensors, and they verify
\begin{eqnarray}
&2\boldsymbol{\sigma}^{\mu\nu}U_{\mu}S_{\nu}\boldsymbol{\pi}\psi\!+\!U^{2}\psi=0\\
&i\Theta S_{\mu}\boldsymbol{\gamma}^{\mu}\psi
\!+\!\Phi S_{\mu}\boldsymbol{\gamma}^{\mu}\boldsymbol{\pi}\psi\!+\!U^{2}\psi=0
\end{eqnarray}
and
\begin{eqnarray}
&\Sigma^{ab}\!=\!-\frac{1}{2}\varepsilon^{abij}M_{ij}\\
&M^{ab}\!=\!\frac{1}{2}\varepsilon^{abij}\Sigma_{ij}
\end{eqnarray}
with
\begin{eqnarray}
&M_{ab}\Phi\!-\!\Sigma_{ab}\Theta\!=\!U^{j}S^{k}\varepsilon_{jkab}\label{A1}\\
&M_{ab}\Theta\!+\!\Sigma_{ab}\Phi\!=\!U_{[a}S_{b]}\label{A2}
\end{eqnarray}
alongside to
\begin{eqnarray}
&M_{ik}U^{i}\!=\!\Theta S_{k}\label{P1}\\
&\Sigma_{ik}U^{i}\!=\!\Phi S_{k}\label{L1}\\
&M_{ik}S^{i}\!=\!\Theta U_{k}\label{P2}\\
&\Sigma_{ik}S^{i}\!=\!\Phi U_{k}\label{L2}
\end{eqnarray}
and with the orthogonality relations
\begin{eqnarray}
&\frac{1}{2}M_{ab}M^{ab}\!=\!-\frac{1}{2}\Sigma_{ab}\Sigma^{ab}\!=\!\Phi^{2}\!-\!\Theta^{2}
\label{norm2}\\
&\frac{1}{2}M_{ab}\Sigma^{ab}\!=\!-2\Theta\Phi
\label{orthogonal2}
\end{eqnarray}
and
\begin{eqnarray}
&U_{a}U^{a}\!=\!-S_{a}S^{a}\!=\!\Theta^{2}\!+\!\Phi^{2}\label{norm1}\\
&U_{a}S^{a}\!=\!0\label{orthogonal1}
\end{eqnarray}
as it is straightforward to demonstrate, called Fierz identities, and which will turn out to have a great importance in the classification of spinors we will do in the following.

From the metric, we define the symmetric connection as usual with $\Lambda^{\sigma}_{\alpha\nu}$ from which, with the tetrads, we define the spin connection $\Omega_{db\pi}\!=\!\eta_{ad}\xi^{\nu}_{b}\xi^{a}_{\sigma}(\Lambda^{\sigma}_{\nu\pi}\!-\!\xi^{\sigma}_{i}\partial_{\pi}\xi_{\nu}^{i})$ which turns out to verify $\Omega_{ab\pi}\!=\!-\Omega_{ba\pi}$ in general. The spinorial connection can always be written according to the form
\begin{eqnarray}
&\boldsymbol{\Omega}_{\mu}
=\frac{1}{2}\Omega^{ab}_{\phantom{ab}\mu}\boldsymbol{\sigma}_{ab}
\!+\!iqA_{\mu}\mathbb{I}\!+\!pC_{\mu}\mathbb{I}\label{spinorialconnection}
\end{eqnarray}
in terms of two real vectors still totally general.

To see this, compute the spinorial covariant derivatives of the gamma matrices and exploit their constancy to get the relation $\boldsymbol{\gamma}^{c}\Omega_{ca\mu}
\!-\![\boldsymbol{\Omega}_{\mu},\boldsymbol{\gamma}_{a}]\!=\!0$ and assume the general expression 
$\boldsymbol{\Omega}_{\mu}\!=\!a\Omega_{ac\mu}\boldsymbol{\sigma}^{ac}\!+\!\boldsymbol{A}_{\mu}$ to plug into it. Once this is done and using (\ref{commgamma}) one gets that $a\!=\!1/2$ and that the vectorial matrix must verify $[\boldsymbol{A}_{\mu},\boldsymbol{\gamma}_{s}]\!=\!0$ which tells that it has to commute with all gamma matrices. Because the gamma matrices and their products generate the space of complex $4\!\times\!4$ matrices, then $\boldsymbol{A}_{\mu}$ commutes with every matrix, and this implies that in its most general form it is proportional to the identity matrix. The proportionality factor is in general complex and therefore it is given with the form
$\boldsymbol{A}_{\mu}\!=\!(iqA_{\mu}\!+\!pC_{\mu})\mathbb{I}$ in terms of two real vectors that are so far still undetermined. However, it is easy to observe that $A_{\mu}$ is the gauge field arising from a unitary phase transformation with charge $q$ and that $C_{\mu}$ is the field arising from conformal transformations $\sigma$ of weight $p$ in general. This demonstrates that the spinor connection is written as in (\ref{spinorialconnection}) in the most general way. Therefore, the most general spinor connection naturally contains a term describing the structure of the space-time, another term describing structure of the gauge potential and one term describing the scaling properties of the space-time.

For the dynamics, we will be focusing on the vectorial field equations given by
\begin{eqnarray}
&\nabla_{\sigma}(\partial V)^{\sigma\mu}\!+\!M^{2}V^{\mu}\!=\!\Gamma^{\mu}\label{F}
\end{eqnarray}
with $(\partial V)_{\sigma\mu}\!=\!\partial_{\sigma}V_{\mu}\!-\!\partial_{\mu}V_{\sigma}$ as usually done. We will also be interested in the spinor differential field equations
\begin{eqnarray}
&i\boldsymbol{\gamma}^{\mu}\boldsymbol{\nabla}_{\mu}\psi
\!-\!XW_{\mu}\boldsymbol{\gamma}^{\mu}\boldsymbol{\pi}\psi\!-\!m\psi\!=\!0\label{D}
\end{eqnarray}
with $X$ the spinor-torsion coupling constant and $W^{\mu}$ the torsion axial-vector added for generality. Notice that for discrete transformations $\psi\!\rightarrow\!\boldsymbol{\pi}\psi$ and $m\!\rightarrow\!-m$ equations (\ref{D}) are invariant. Finally we also notice that conformal invariance would require $m\!=\!0$ to hold identically.

It is important to notice that if we multiply on the left by the $\mathbb{I}$, $\boldsymbol{\gamma}^{a}$, $\boldsymbol{\sigma}^{ab}$, $\boldsymbol{\gamma}^{a}\boldsymbol{\pi}$, $\boldsymbol{\pi}$ and by the complex conjugate spinor field, splitting real and imaginary parts gives
\begin{eqnarray}
&\frac{i}{2}(\overline{\psi}\boldsymbol{\gamma}^{\mu}\boldsymbol{\nabla}_{\mu}\psi
\!-\!\boldsymbol{\nabla}_{\mu}\overline{\psi}\boldsymbol{\gamma}^{\mu}\psi)
\!-\!XW_{\sigma}S^{\sigma}\!-\!m\Phi\!=\!0\\
&\nabla_{\mu}U^{\mu}\!=\!0
\end{eqnarray}
\begin{eqnarray}
&\frac{i}{2}(\overline{\psi}\boldsymbol{\gamma}^{\mu}\boldsymbol{\pi}\boldsymbol{\nabla}_{\mu}\psi
\!-\!\boldsymbol{\nabla}_{\mu}\overline{\psi}\boldsymbol{\gamma}^{\mu}\boldsymbol{\pi}\psi)
\!-\!XW_{\sigma}U^{\sigma}\!=\!0\\
&\nabla_{\mu}S^{\mu}\!-\!2m\Theta\!=\!0
\end{eqnarray}
\begin{eqnarray}
\nonumber
&i(\overline{\psi}\boldsymbol{\nabla}^{\alpha}\psi
\!-\!\boldsymbol{\nabla}^{\alpha}\overline{\psi}\psi)
\!-\!\nabla_{\mu}M^{\mu\alpha}-\\
&-XW_{\sigma}M_{\mu\nu}\varepsilon^{\mu\nu\sigma\alpha}\!-\!2mU^{\alpha}\!=\!0
\label{vr}\\
\nonumber
&\nabla_{\alpha}\Phi
\!-\!2(\overline{\psi}\boldsymbol{\sigma}_{\mu\alpha}\!\boldsymbol{\nabla}^{\mu}\psi
\!-\!\boldsymbol{\nabla}^{\mu}\overline{\psi}\boldsymbol{\sigma}_{\mu\alpha}\psi)+\\
&+2X\Theta W_{\alpha}\!=\!0\label{vi}
\end{eqnarray}
\begin{eqnarray}
\nonumber
&\nabla_{\nu}\Theta\!-\!
2i(\overline{\psi}\boldsymbol{\sigma}_{\mu\nu}\boldsymbol{\pi}\boldsymbol{\nabla}^{\mu}\psi\!-\!
\boldsymbol{\nabla}^{\mu}\overline{\psi}\boldsymbol{\sigma}_{\mu\nu}\boldsymbol{\pi}\psi)-\\
&-2X\Phi W_{\nu}\!+\!2mS_{\nu}\!=\!0\label{ar}\\
\nonumber
&(\boldsymbol{\nabla}_{\alpha}\overline{\psi}\boldsymbol{\pi}\psi
\!-\!\overline{\psi}\boldsymbol{\pi}\boldsymbol{\nabla}_{\alpha}\psi)
\!-\!\frac{1}{2}\nabla^{\mu}M^{\rho\sigma}\varepsilon_{\rho\sigma\mu\alpha}+\\
&+2XW^{\mu}M_{\mu\alpha}\!=\!0\label{ai}
\end{eqnarray}
\begin{eqnarray}
\nonumber
&\nabla^{\mu}S^{\rho}\varepsilon_{\mu\rho\alpha\nu}
\!+\!i(\overline{\psi}\boldsymbol{\gamma}_{[\alpha}\!\boldsymbol{\nabla}_{\nu]}\psi
\!-\!\!\boldsymbol{\nabla}_{[\nu}\overline{\psi}\boldsymbol{\gamma}_{\alpha]}\psi)+\\
&+2XW_{[\alpha}S_{\nu]}\!=\!0\\
\nonumber
&\nabla^{[\alpha}U^{\nu]}\!+\!i\varepsilon^{\alpha\nu\mu\rho}
(\overline{\psi}\boldsymbol{\gamma}_{\rho}\boldsymbol{\pi}\!\boldsymbol{\nabla}_{\mu}\psi\!-\!\!
\boldsymbol{\nabla}_{\mu}\overline{\psi}\boldsymbol{\gamma}_{\rho}\boldsymbol{\pi}\psi)-\\
&-2XW_{\sigma}U_{\rho}\varepsilon^{\alpha\nu\sigma\rho}\!-\!2mM^{\alpha\nu}\!=\!0
\end{eqnarray}
as it is direct to see, and called Gordon decompositions.
\section{The Polar Forms}
In treating spinors, a possible way to simplify the study is to write them in polar form, where each component is expressed as a module times a unitary phase, while still respecting their spinorial covariance. In doing this, a first advantage is that the spinor formalism is converted into one in which all relevant quantities are real tensors, but more importantly, of all these quantities it is possible to tell apart the physical degrees of freedom from all other non-physical components. It may then be interesting to ask whether a similar decomposition is possible also for fields that are not spinors, as for instance for the case of vectors. Although vectors are already real, it would still be interesting to write them in a way that isolates all the degrees of freedom from all the non-physical components.

We will recall the polar form of spinors as in \cite{Fabbri:2016msm, Fabbri:2016laz} and \cite{Fabbri:2017pwp}. Then we apply the same argument for vectors.

\subsection{Spinor Fields}
We split two cases, when both $\Theta$ and $\Phi$ are identically equal to zero and when at least one of them is not.

In the latter case, called regular, identity (\ref{norm1}) tells that $U^{a}$ is time-like, and so we can always perform up to three boosts in order to bring its spatial components to vanish identically. Then it is in general always possible to take advantage of up to two rotations to bring the space part of $S^{a}$ aligned along the third axis, eventually employing the last rotation to isolate a unitary phase. Because the unitary phase has already been isolated, gauge transformations have no effect that cannot be already attributed to a rotation around the third axis. However, conformal transformations $\sigma$ can also be used for the normalization of the overall spinor field. If we call the set of all Lorentz transformations times the gauge phase collectively $\boldsymbol{S}^{-1}$ then it is easy to see that the most general spinorial field can always be written in terms of the polar form
\begin{eqnarray}
&\!\!\psi\!=\!\phi e^{-\frac{i}{2}\beta\boldsymbol{\pi}}
\boldsymbol{S}\left(\!\begin{tabular}{c}
$1$\\
$0$\\
$1$\\
$0$
\end{tabular}\!\right)
\label{regular}
\end{eqnarray}
with $\beta$ and $\phi$ called Yvon-Takabayashi angle and module, respectively. With conformal transformations $\sigma$ we may pick the scale factor to be $\sigma\!=\!\phi^{2/3}$ and by including also the scale factor within $\boldsymbol{S}^{-1}$ then we would reduce to
\begin{eqnarray}
&\!\!\psi\!=\!e^{-\frac{i}{2}\beta\boldsymbol{\pi}}
\boldsymbol{S}\left(\!\begin{tabular}{c}
$1$\\
$0$\\
$1$\\
$0$
\end{tabular}\!\right)
\label{regularnormal}
\end{eqnarray}
in terms of the Yvon-Takabayashi angle alone. It is very important to remark that in the most general case these spinors are fields, that is point-dependent, and therefore any transformation bringing them in polar form must be a local transformation, hence point-dependent. When we will compute the derivatives, the local parameters in the transformations will give rise to connection-like terms.

For the case of the regular spinorial fields, we have that the two antisymmetric tensorial bi-linear quantities given by $\Sigma^{ab}$ and $M^{ab}$ can always be written according to
\begin{eqnarray}
&\Sigma^{ab}\!=\!2\phi^{2}(\cos{\beta}u^{[a}s^{b]}\!-\!\sin{\beta}u_{j}s_{k}\varepsilon^{jkab})\\
&M^{ab}\!=\!2\phi^{2}(\cos{\beta}u_{j}s_{k}\varepsilon^{jkab}\!+\!\sin{\beta}u^{[a}s^{b]})
\end{eqnarray}
in terms of the two vectors defined as
\begin{eqnarray}
&S^{a}\!=\!2\phi^{2}s^{a}\\
&U^{a}\!=\!2\phi^{2}u^{a}
\end{eqnarray}
and with the scalars
\begin{eqnarray}
&\Theta\!=\!2\phi^{2}\sin{\beta}\\
&\Phi\!=\!2\phi^{2}\cos{\beta}
\end{eqnarray}
having the consequence that all the Fierz identities reduce to be trivial with the exception of the conditions
\begin{eqnarray}
&2\boldsymbol{\sigma}^{\mu\nu}u_{\mu}s_{\nu}\boldsymbol{\pi}\psi\!+\!\psi=0\label{aux1}\\
&i\sin{\beta}s_{\mu}\boldsymbol{\gamma}^{\mu}\psi
\!+\!\cos{\beta}s_{\mu}\boldsymbol{\gamma}^{\mu}\boldsymbol{\pi}\psi\!+\!\psi=0\label{aux2}
\end{eqnarray}
and the relationships
\begin{eqnarray}
&u_{a}u^{a}\!=\!-s_{a}s^{a}\!=\!1\\
&u_{a}s^{a}\!=\!0
\end{eqnarray}
as it is easy to see. These last expressions also show that the two vectors are not free leaving the Yvon-Takabayashi angle and the module as the only two physical degrees of freedom, and that Yvon-Takabayashi angle and module are a pseudo-scalar and a scalar. Notice that the spinorial field in its polar form is fixed up to reversals of the third axis and up to discrete transformations $\beta\!\rightarrow\!\beta\!+\!\pi$ as it is quite clear. We remark that the Yvon-Takabayashi angle would remain the only physical degree of freedom, in case the module is normalized by conformal transformations.

For the instance of regular spinor fields, recalling that we adopted the convention of including the gauge transformation within the $\boldsymbol{S}$ matrix, we have that
\begin{eqnarray}
&\boldsymbol{S}\partial_{\mu}\boldsymbol{S}^{-1}\!=\!i\partial_{\mu}\alpha\mathbb{I}
\!+\!\frac{1}{2}\partial_{\mu}\theta_{ij}\boldsymbol{\sigma}^{ij}\label{spintrans}
\end{eqnarray}
where $\alpha$ is the generic complex phase and $\theta_{ij}\!=\!-\theta_{ji}$ are the six parameters of the Lorentz group. So we can define
\begin{eqnarray}
&\partial_{\mu}\alpha\!-\!qA_{\mu}\!\equiv\!P_{\mu}\label{P}\\
&\partial_{\mu}\theta_{ij}\!-\!\Omega_{ij\mu}\!\equiv\!R_{ij\mu}\label{R}
\end{eqnarray}
which can be proven to be tensors and invariant under a gauge transformation simultaneously. Phase and parameters do not alter the information within gauge potential and spin connection, but the non-physical components of spinors encoded by phase and parameters combine with the non-covariant properties of gauge potential and spin connection in order to ensure covariance of $P_{\mu}$ and $R_{ij\mu}$ eventually. For this reason, expressions (\ref{P}, \ref{R}) are called gauge-invariant vector momentum and tensorial connection, and we may simply refer to them altogether as tensorial connections. With them we can finally compute
\begin{eqnarray}
&\boldsymbol{\nabla}_{\mu}\psi\!=\!(-\frac{i}{2}\nabla_{\mu}\beta\boldsymbol{\pi}
\!+\!\nabla_{\mu}\ln{\phi}\mathbb{I}
\!-\!iP_{\mu}\mathbb{I}\!-\!\frac{1}{2}R_{ij\mu}\boldsymbol{\sigma}^{ij})\psi
\label{decspinder}
\end{eqnarray}
as spinorial covariant derivative. When conformal transformations $\sigma$ are included within the $\boldsymbol{S}$ matrix
\begin{eqnarray}
&\boldsymbol{S}\partial_{\mu}\boldsymbol{S}^{-1}\!=\!i\partial_{\mu}\alpha\mathbb{I}
\!+\!\frac{1}{2}\partial_{\mu}\theta_{ij}\boldsymbol{\sigma}^{ij}
\!-\!3/2\partial_{\mu}\ln{\sigma}
\label{spinconftrans}
\end{eqnarray}
where $\sigma$ is the scale factor. With $pC_{\mu}\!=\!-3/2\nabla_{\mu}\ln{\sigma}$ we can still assume the same (\ref{P}, \ref{R}) to compute the spinor covariant derivative. It is then given by
\begin{eqnarray}
&\boldsymbol{\nabla}_{\mu}\psi\!=\!(-\frac{i}{2}\nabla_{\mu}\beta\boldsymbol{\pi}
\!-\!iP_{\mu}\mathbb{I}\!-\!\frac{1}{2}R_{ij\mu}\boldsymbol{\sigma}^{ij})\psi
\label{decspinconfder}
\end{eqnarray}
as spinorial conformal covariant derivative. Either way
\begin{eqnarray}
&\nabla_{\mu}s_{i}\!=\!R_{ji\mu}s^{j}\label{ds}\\
&\nabla_{\mu}u_{i}\!=\!R_{ji\mu}u^{j}\label{du}
\end{eqnarray}
are general identities. From the commutators we have
\begin{eqnarray}
\!\!\!\!&qF_{\mu\nu}\!=\!-(\nabla_{\mu}P_{\nu}\!-\!\nabla_{\nu}P_{\mu})\label{Maxwell}\\
&\!\!\!\!\!\!\!\!R^{i}_{\phantom{i}j\mu\nu}\!=\!-(\nabla_{\mu}R^{i}_{\phantom{i}j\nu}
\!-\!\!\nabla_{\nu}R^{i}_{\phantom{i}j\mu}
\!\!+\!R^{i}_{\phantom{i}k\mu}R^{k}_{\phantom{k}j\nu}
\!-\!R^{i}_{\phantom{i}k\nu}R^{k}_{\phantom{k}j\mu})\label{Riemann}
\end{eqnarray}
which tell us that the parameters defined in (\ref{spintrans}) do not generate any curvature tensor of their own, and therefore gauge potential and spin connection transfer only physical information into $F_{\mu\nu}$ and $R^{i}_{\phantom{i}j\mu\nu}$ respectively. Objects (\ref{Maxwell}, \ref{Riemann}) are the Maxwell and Riemann curvature, and as such they encode electrodynamic and gravitational information as usual. However, absence of physical fields, and that is vanishing of these curvatures, does not necessarily imply the vanishing of gauge-invariant vector momentum and tensorial connection, which can still be different from zero in general. Hence, the tensorial connections encode information of covariant type about inertial and gravitational accelerations, or about inertial acceleration alone when their curvatures are identically equal to zero \cite{Fabbri:2018crr}.

Plugging into (\ref{D}) the polar form gives
\begin{eqnarray}
&B_{\mu}\!-\!2P^{\iota}u_{[\iota}s_{\mu]}\!+\!(\nabla\beta\!-\!2XW)_{\mu}
\!+\!2s_{\mu}m\cos{\beta}\!=\!0\label{dep1}\\
&R_{\mu}\!-\!2P^{\rho}u^{\nu}s^{\alpha}\varepsilon_{\mu\rho\nu\alpha}\!+\!2s_{\mu}m\sin{\beta}
\!+\!\nabla_{\mu}\ln{\phi^{2}}\!=\!0\label{dep2}
\end{eqnarray}
with $R_{\mu a}^{\phantom{\mu a}a}\!=\!R_{\mu}$ and $\frac{1}{2}\varepsilon_{\mu\alpha\nu\iota}R^{\alpha\nu\iota}\!=\!B_{\mu}$ and which can be proven to imply the spinor differential field equations, so that (\ref{D}) and (\ref{dep1}, \ref{dep2}) are fully equivalent. To prove this, have the polar form, together with all bi-linear quantities, plugged into decompositions (\ref{vi}, \ref{ar}), getting
\begin{eqnarray}
\nonumber
&\frac{1}{2}\nabla_{\alpha}\ln{\phi^{2}}\cos{\beta}
\!-\!(\frac{1}{2}\nabla_{\alpha}\beta\!-\!XW_{\alpha})\sin{\beta}+\\
\nonumber
&+P^{\mu}(u^{\rho}s^{\sigma}\varepsilon_{\rho\sigma\mu\alpha}\cos{\beta}
\!+\!u_{[\mu}s_{\alpha]}\sin{\beta})+\\
&+\frac{1}{2}R_{\alpha\mu}^{\phantom{\alpha\mu}\mu}\cos{\beta}
\!+\!\frac{1}{4}R^{\rho\sigma\mu}\varepsilon_{\rho\sigma\mu\alpha}\sin{\beta}\!=\!0\\
\nonumber
&\frac{1}{2}\nabla_{\nu}\ln{\phi^{2}}\sin{\beta}
\!+\!(\frac{1}{2}\nabla_{\nu}\beta\!-\!XW_{\nu})\cos{\beta}+\\
\nonumber
&+P^{\mu}(u^{\rho}s^{\sigma}\varepsilon_{\rho\sigma\mu\nu}\sin{\beta}\!-\!u_{[\mu}s_{\nu]}\cos{\beta})-\\
&-\frac{1}{4}R^{\rho\sigma\mu}\varepsilon_{\rho\sigma\mu\nu}\cos{\beta}
\!+\!\frac{1}{2}R_{\nu\mu}^{\phantom{\nu\mu}\mu}\sin{\beta}\!+\!ms_{\nu}\!=\!0
\end{eqnarray}
which can be diagonalized into
\begin{eqnarray}
\nonumber
&\frac{1}{2}\varepsilon_{\mu\alpha\nu\iota}R^{\alpha\nu\iota}
\!-\!2P^{\iota}u_{[\iota}s_{\mu]}-\\
&-2XW_{\mu}\!+\!\nabla_{\mu}\beta\!+\!2s_{\mu}m\cos{\beta}\!=\!0\\
\nonumber
&R_{\mu a}^{\phantom{\mu a}a}
\!-\!2P^{\rho}u^{\nu}s^{\alpha}\varepsilon_{\mu\rho\nu\alpha}+\\
&+2s_{\mu}m\sin{\beta}\!+\!\nabla_{\mu}\ln{\phi^{2}}\!=\!0
\end{eqnarray}
while the converse is proven by considering these as well as the general identities (\ref{aux1}, \ref{aux2}) to work out
\begin{eqnarray}
\nonumber
&i\boldsymbol{\gamma}^{\mu}\boldsymbol{\nabla}_{\mu}\psi
\!-\!XW_{\sigma}\boldsymbol{\gamma}^{\sigma}\boldsymbol{\pi}\psi\!-\!m\psi=\\
\nonumber
&=[i\boldsymbol{\gamma}^{\mu}P^{\rho}u^{\nu}s^{\alpha}\varepsilon_{\mu\rho\nu\alpha}+\\
\nonumber
&+P^{\iota}u_{[\iota}s_{\mu]}\boldsymbol{\gamma}^{\mu}\boldsymbol{\pi}
\!+\!P_{\mu}\boldsymbol{\gamma}^{\mu}-\\
&-is_{\mu}\boldsymbol{\gamma}^{\mu}m\sin{\beta}
\!-\!s_{\mu}\boldsymbol{\gamma}^{\mu}\boldsymbol{\pi}m\cos{\beta}\!-\!m\mathbb{I}]\psi\!=\!0
\end{eqnarray}
showing that the spinorial differential field equations are valid in polar form and therefore in general. Notice that the spinor equation, being $4$ complex equations, amounts to $8$ real equations, which are as many as those provided by the $2$ vector equations. These two vectorial equations give all derivatives of the two physical degrees of freedom, given by Yvon-Takabayashi angle and module. It is also important to remark that $\beta\!\!\rightarrow\!\!\beta+\pi$ requires that $m\!\rightarrow\!-m$ in order for (\ref{dep1}, \ref{dep2}) to be invariant. We notice also that for conformal invariance the module must be normalized to unity and $m\!=\!0$ with the consequence that (\ref{dep1}) would remain one field equation while (\ref{dep2}) would convert into one constraint. We observe that in (\ref{dep1}, \ref{dep2}) the tensorial connections act as some tension of the space-time which may have effects on Yvon-Takabayashi angle and module, even if in some cases it is not rich enough to give rise to either electrodynamics or gravitation at all \cite{Fabbri:2019kfr}.

In the complementary case, called singular, by writing $M_{0K}\!=\!E_{K}$ and $M_{IJ}\!=\!\varepsilon_{IJK}B^{K}$ identities (\ref{norm2}, \ref{orthogonal2}) convert into $E^{2}\!=\!B^{2}$ and $\vec{E}\!\cdot\!\vec{B}\!=\!0$ so that we can always employ boosts and rotations to bring the two three-dimensional vectors aligned along given axes. By collecting the gauge phase and Lorentz transformations into a single $\boldsymbol{S}^{-1}$ we can write the most general spinorial field in this case as
\begin{eqnarray}
&\!\!\psi\!=\!\boldsymbol{S}\left(\!\begin{tabular}{c}
$\cos{\frac{\theta}{2}}$\\
$0$\\
$0$\\
$\sin{\frac{\theta}{2}}$
\end{tabular}\!\right)
\label{singular1}
\end{eqnarray}
with $\theta$ a generic angle still unspecified. Notice that there is no way to factor it out from the structure of the spinor.

For singular spinors, the only relation we have is
\begin{eqnarray}
&S^{a}\!=\!-\cos{\theta}U^{a}
\end{eqnarray}
showing that $\theta$ is a function that describes the projection of the spin axial-vector onto the velocity vector, and thus the $\theta$ angle can be interpreted as the helicity. It may be interesting to recall that this type of spinor is named flag di-pole spinor, when subject to the constraint $\theta\!=\!\pm\pi/2$ it is the Majorana spinor and $\theta\!=\!0,\pi$ it is the Weyl spinor.

For singular spinor fields, we still have that (\ref{spintrans}) holds, along with (\ref{P}, \ref{R}), so that we can still write the spinorial covariant derivative according to
\begin{eqnarray}
&\boldsymbol{\nabla}_{\mu}\psi\!=\!(-\frac{i}{2}\nabla_{\mu}\theta 
\boldsymbol{S}\boldsymbol{\gamma}^{2}\boldsymbol{S}^{-1}\boldsymbol{\pi}
\!-\!iP_{\mu}\mathbb{I}\!-\!\frac{1}{2}R_{ij\mu}\boldsymbol{\sigma}^{ij})\psi
\end{eqnarray}
but we see that in it a strange $\boldsymbol{S}\boldsymbol{\gamma}^{2}\boldsymbol{S}^{-1}$ term has appeared and which seems to break Lorentz symmetry by selecting the second axis as privileged. We remark that this $\theta$ angle is supposed to be a physical degree of freedom for flag di-pole spinors, and this adds another weird feature because as such it can freely move from all possible values ranging from those defining a Majorana spinor to those defining a Weyl spinor. Thus it seems always possible for a Weyl spinor to mutate into a Majorana spinor, although such an occurrence has never been observed to happen. A way out of this situation may be to assume $\theta$ constant, which would remove the frame-dependent term in the derivative while forbidding a Weyl spinor and a Majorana spinor to transmute into one another. However, in this case, a flag di-pole remains without any physical degree of freedom.

So, employing the Gordon decompositions to get polar field equations would result into constraints.

\subsection{Vector Fields}
Compared to the case of spinors, where many bi-linear quantities and their identities helped in classifying fields, in the case of vector we lack all this, so the classification will turn out to be different, and instead of a classification of different types of spinors, it will results into a splitting into different behaviours of the vector field. We thus split three cases, given by $V^{2}\!>\!0$, $V^{2}\!=\!0$ or $V^{2}\!<\!0$ in general.

In the case $V^{2}\!>\!0$ we can recover the same derivation done above, because since $V^{a}$ is time-like then it is always possible to perform up to three boosts to bring its spatial components to vanish. Collecting these transformations into $\Lambda^{-1}$ we can always write the vector in polar form
\begin{eqnarray}
&V\!=\!\phi\Lambda\left(\!\begin{tabular}{c}
$1$\\
$0$\\
$0$\\
$0$
\end{tabular}\!\right)
\label{pos}
\end{eqnarray}
with $\phi$ as module, which is proven to be scalar, and it is the only physical degree of freedom in general. Conformal transformations remove the module, leaving such a vector with no physical degree of freedom. In the case $V^{2}\!=\!0$ it is always possible to perform rotations to align the space part of the vector along the third axis. Because for such case the zeroth and the third components are equal, they can be factored and then a boost along the third axis can be used to normalize them to unity. So in this case
\begin{eqnarray}
&V\!=\!\Lambda\left(\!\begin{tabular}{c}
$1$\\
$0$\\
$0$\\
$1$
\end{tabular}\!\right)
\label{nul}
\end{eqnarray}
with no physical degree of freedom. In the case $V^{2}\!<\!0$ it is always possible to perform up to three boosts to bring the time component to zero and some rotations to have the space part aligned along the third axis. By collecting the transformations into $\Lambda^{-1}$ the vector in polar form is
\begin{eqnarray}
&V\!=\!\phi\Lambda\left(\!\begin{tabular}{c}
$0$\\
$0$\\
$0$\\
$1$
\end{tabular}\!\right)
\label{neg}
\end{eqnarray}
with $\phi$ as the module, still a scalar, and the only physical degree of freedom. A conformal transformation removes the module by normalizing it to unity, and therefore there is no physical degree of freedom left in the vector.

In the following, it will be simpler to define
\begin{eqnarray}
&V^{a}\!=\!\phi v^{a}
\end{eqnarray}
in terms of $v^{b}$ such that $v^{2}\!=\!1$, $v^{2}\!=\!0$ or $v^{2}\!=\!-1$ for the three cases above since it is just the normalized vector.

Then in any of these cases we have that we can write
\begin{eqnarray}
&(\Lambda)^{i}_{\phantom{i}k}\partial_{\mu}(\Lambda^{-1})^{k}_{\phantom{k}j}
\!=\!\partial_{\mu}\theta^{i}_{\phantom{i}j}
\end{eqnarray}
where $\theta^{i}_{\phantom{i}j}\!=\!\theta^{ik}\eta_{kj}$ are the six parameters of the Lorentz group as above. Therefore we can define
\begin{eqnarray}
&\partial_{\mu}\theta^{i}_{\phantom{i}j}\!-\!\Omega^{i}_{\phantom{i}j\mu}
\!\equiv\!R^{i}_{\phantom{i}j\mu}
\end{eqnarray}
again as above. So as before (\ref{R}) is the expression of the tensorial connection. With it we can compute
\begin{eqnarray}
&\nabla_{\mu}V^{a}\!=\!(\delta^{a}_{b}\nabla_{\mu}\ln{\phi}\!-\!R^{a}_{\phantom{a}b\mu})V^{b}
\label{decvecder}
\end{eqnarray}
as covariant derivative. Then we have
\begin{eqnarray}
&\nabla_{\mu}v_{a}\!=\!R_{ba\mu}v^{b}\label{dv}
\end{eqnarray}
as general identities. The commutator still gives
\begin{eqnarray}
&\!\!\!\!\!\!\!\!\!\!\!R^{\alpha}_{\phantom{\alpha}\rho\mu\nu}
\!\!=\!-(\nabla_{\mu}R^{\alpha}_{\phantom{\alpha}\rho\nu}
\!-\!\!\nabla_{\nu}R^{\alpha}_{\phantom{\alpha}\rho\mu}
\!\!+\!R^{\alpha}_{\phantom{\alpha}k\mu}R^{k}_{\phantom{k}\rho\nu}
\!\!-\!\!R^{\alpha}_{\phantom{\alpha}k\nu}R^{k}_{\phantom{k}\rho\mu})
\end{eqnarray}
as the expression for the Riemann curvature in general.

When in (\ref{F}) we plug the polar form we get
\begin{eqnarray}
\nonumber
&(g^{\alpha\nu}\nabla^{2}\phi\!-\!\nabla^{\nu}\nabla^{\alpha}\phi-\\
\nonumber
&-R^{\nu}\nabla^{\alpha}\phi\!+\!R^{\nu\alpha\sigma}\nabla_{\sigma}\phi
\!+\!R^{\nu[\alpha\sigma]}\nabla_{\sigma}\phi+\\
&+\nabla_{\sigma}R^{\nu[\alpha\sigma]}\phi
\!+\!R^{\sigma[\alpha\pi]}R^{\nu}_{\phantom{\nu}\sigma\pi}\phi
\!+\!M^{2}g^{\alpha\nu}\phi)v_{\nu}\!=\!\Gamma^{\alpha}
\end{eqnarray}
as the polar form of the field equations for vector fields.
\section{Non-Locality}
Thus far, we have worked out the details of the general theory. We have seen that for spinor fields as well as for vector fields, it is possible to employ active local Lorentz transformations, possibly complemented by gauge shifts, and conformal scalings, in order to write the most general of these fields in special forms, called polar forms. These polar forms have the advantage of isolating the physical degrees of freedom in order to keep them separated from the non-physical components, which can be used together with the connections to define objects that still contain the same information of connections but which are also proven to be tensors. Such tensorial connections, defined by (\ref{P}, \ref{R}), contain information about a covariant type of accelerations. They may remain non-zero even if their curvatures vanish identically. In such a case, they contain information about a covariant type of sourceless accelerations because no physical field is present. Just the same, there are physical effects in these covariant sourceless accelerations, which can be pictured as some tension of the space-time that may have effects even if not rich enough to generate curvatures that can couple to a source \cite{Fabbri:2019kfr}.

A specific case of such effects can be seen after we have manipulated field equations (\ref{D}) and (\ref{F}) written in their polar form. To this purpose, we consider (\ref{D}) and apply a second time the derivative so to go at the second-order derivative, where the polar form gives
\begin{eqnarray}
&\!\!\!\!\nabla_{\mu}(\phi^{2}\nabla^{\mu}\beta/2)
\!-\!\frac{1}{2}(\nabla_{\mu}K^{\mu}\!+\!K_{\mu}G^{\mu})\phi^{2}\!=\!0\label{cont}\\
&\!\!\!\!\!\!\!\!\left|\nabla\beta/2\right|^{2}\!\!-\!m^{2}\!-\!\nabla^{2}\phi/\phi
\!+\!\frac{1}{2}(\nabla_{\mu}G^{\mu}\!+\!\frac{1}{2}G^{2}\!-\!\frac{1}{2}K^{2})\!=\!0\label{HJ}
\end{eqnarray}
with vector potential $G_{\mu}\!=\!-R_{\mu}\!+\!2P^{\rho}u^{\nu}s^{\alpha}\varepsilon_{\mu\rho\nu\alpha}$ and with axial-vector potential $K_{\mu}\!=\!2XW_{\mu}\!-\!B_{\mu}\!+\!2P^{\iota}u_{[\iota}s_{\mu]}$ defined for simplicity. We remark that the field equation for the Yvon-Takabayashi angle (\ref{cont}) is in the form of a continuity equation while the field equation for the module (\ref{HJ}) has the structure of the Hamilton-Jacobi equation. We could do the same procedure for (\ref{F}) by extracting its primary constraint $M^{2}\nabla V\!=\!\nabla \Gamma$ and substituting it back into the original field equation (\ref{F}) therefore getting
\begin{eqnarray}
&\nabla^{2}V^{\mu}\!-\!R^{\mu\rho}V_{\rho}
\!+\!M^{2}V^{\mu}\!=\!\Gamma^{\mu}\!+\!M^{-2}\nabla^{\mu}\nabla \Gamma
\end{eqnarray}
which in polar form is
\begin{eqnarray}
\nonumber
&(g^{\sigma\alpha}\nabla^{2}\phi\!-\!2R^{\sigma\alpha\mu}\nabla_{\mu}\phi
\!-\!\nabla_{\mu}R^{\sigma\alpha\mu}\phi
\!-\!R^{\sigma}_{\phantom{\sigma}\nu\mu}R^{\alpha\nu\mu}\phi-\\
&-R^{\sigma\alpha}\phi\!+\!M^{2}g^{\sigma\alpha}\phi)v_{\alpha}
\!=\!\Gamma^{\sigma}\!+\!M^{-2}\nabla^{\sigma}\nabla\Gamma
\end{eqnarray}
in general. In particular, contracting with the vector
\begin{eqnarray}
\nonumber
&\nabla^{2}\phi\!+\!(v_{\sigma}R^{\sigma}_{\phantom{\sigma}\nu\mu}v_{\alpha}R^{\alpha\nu\mu}
\!+\!v_{\sigma}v_{\alpha}R^{\sigma\alpha}+\\
&+M^{2})\phi\!=\!-(v_{\sigma}\Gamma^{\sigma}\!+\!M^{-2}v_{\sigma}\nabla^{\sigma}\nabla\Gamma)
\label{L}
\end{eqnarray}
which is an equation that is similar to (\ref{HJ}).

By taking (\ref{HJ}) and (\ref{L}) in the free case and also setting the Yvon-Takabayashi angle to zero we get that
\begin{eqnarray}
&\!\!-\nabla^{2}\phi/\phi\!=\!m^{2}
\!+\!\frac{1}{2}(\nabla R\!-\!\frac{1}{2}R^{2}\!+\!\frac{1}{2}B^{2})\\
&\!\!-\nabla^{2}\phi/\phi\!=\!M^{2}
\!+\!v_{\sigma}R^{\sigma}_{\phantom{\sigma}\nu\mu}v_{\alpha}R^{\alpha\nu\mu}
\!+\!v_{\sigma}v_{\alpha}R^{\sigma\alpha}
\end{eqnarray}
isolating the quantum potential $-\nabla^{2}\phi/\phi$ in both expressions. Since the tensorial connection $R_{ij\mu}$ cannot be vanished in general, then there will always be corrections to the mass term, keeping the quantum potential from vanishing. The presence of these non-trivial quantum potentials makes up for the non-local features in the dynamics of elementary particles. While this was known for spinors now we see that the same also occurs for vectors.
\section{Conclusion}
In this paper, we have recalled recent developments on the role of the polar decomposition of spinor fields for the procedure of having the components re-arranged to keep isolated real degrees of freedom from the components due to the non-trivial structure of the tetradic frame, and we have seen that these last can be combined with the spin connection to form quantities that contain the same information of the connection but which are tensors. Such tensorial connection encodes information about a covariant type of acceleration that can be either inertial if the curvature vanishes or inertial and gravitational if the curvature is generally different from zero. Very interestingly, the tensorial connection may remain non-zero even if its curvature vanishes, so there can be effects even for accelerations due to no external source. They can be regarded as tensions of space-time itself. This construction is now very well-known in the case of the spinorial fields.

In this paper we have extended it to the case of vectorial fields. We have found a way to split the real degrees of freedom from the components due to the tetradic frame and we had these last combined with the spin connection to form the tensorial connection exactly as above.

The difference between spinor and vector fields is that the former are complex and thus also the gauge-invariant vector momentum could eventually be defined.

The form of a relativistic version of the quantum potential has been found for the second-order derivative field equations in both of these situations.


\begin{thebibliography}{40}
\bibitem{G}
M.Gasperini, \textit{Theory of Gravitational\\ Interactions} (Springer, 2017).
\bibitem{L}
P.Lounesto, \textit{Clifford Algebras and\\ Spinors} (Cambridge University Press, 2001).
\bibitem{Cavalcanti:2014wia}
R.T.Cavalcanti, ``Classification of Singular Spinor\\ Fields and Other Mass Dimension One Fermions'',\\ \textit{Int.J.Mod.Phys.D}\textbf{23}, 1444002 (2014).
\bibitem{HoffdaSilva:2017waf} 
J.M.Hoff da Silva, R.T.Cavalcanti, ``Revealing how\\ different spinors can be: the Lounesto 
spinor\\ classification'', \textit{Mod.Phys.Lett.A}\textbf{32}, 1730032 (2017).
\bibitem{daSilva:2012wp}
J.M.Hoff da Silva, R.da Rocha, ``Unfolding Physics\\ from the Algebraic Classification of Spinor\\ Fields'', \textit{Phys. Lett. B}\textbf{718}, 1519 (2013).
\bibitem{daRocha:2008we} 
R.da Rocha, J.M.Hoff da Silva, ``ELKO, flagpole and\\ flag-dipole spinor fields, and the instanton Hopf\\ fibration'', \textit{Adv. Appl. Clifford Algebras} \textbf{20}, 847 (2010).
\bibitem{Villalobos:2015xca}
C.H.Coronado Villalobos, J.M.Hoff da Silva, R.da Rocha,\\
``Questing mass dimension $1$ spinor fields'',\\
\textit{Eur.Phys.J.C} \textbf{75}, 266 (2015).
\bibitem{Ablamowicz:2014rpa}
R.Ab{\l}amowicz, I.Gon{\c c}alves, R.da Rocha, ``Bilinear\\ Covariants and Spinor Fields Duality in Quantum\\ Clifford Algebras'', \textit{J. Math. Phys.}\textbf{55}, 103501 (2014).
\bibitem{Vignolo:2011qt}
S.Vignolo, L.Fabbri, R.Cianci,
``Dirac spinors in\\ Bianchi-I f(R)-cosmology with torsion'',\\
\textit{J.Math.Phys.}\textbf{52}, 112502 (2011).
\bibitem{daRocha:2013qhu}
R.da Rocha,L.Fabbri,J.M.Hoff da Silva,R.T.Cavalcanti, J.A.Silva-Neto,
``Flag-Dipole Spinor Fields in ESK Gravities'',
\textit{J.Math.Phys.}\textbf{54},102505(2013).
\bibitem{Ahluwalia:2004sz}
D.V.Ahluwalia, D.Grumiller, ``Dark matter: A Spin\\ 
one half fermion field with mass dimension one?'',\\
\textit{Phys.Rev.D}\textbf{72}, 067701 (2005).
\bibitem{Ahluwalia:2004ab} 
D.V.Ahluwalia, D.Grumiller,
``Spin half fermions with mass dimension one: Theory, phenomenology, and dark matter'',
\textit{JCAP}\textbf{0507}, 012 (2005).
\bibitem{Ahluwalia:2016rwl} 
D.V.Ahluwalia,
``The theory of local mass\\ dimension one fermions of spin one half'',\\
\textit{Adv.Appl.Clifford Algebras}\textbf{27}, 2247 (2017).
\bibitem{Ahluwalia:2016jwz}
D.V.Ahluwalia, ``Evading Weinberg's no-go theorem to\\ construct mass dimension one fermions: Constructing\\ darkness'', \textit{EPL}\textbf{118}, 60001 (2017).
\bibitem{daRocha:2007pz}
R.da Rocha, J.M.Hoff da Silva,
``From Dirac spinor fields to ELKO'',
\textit{J.Math.Phys.}\textbf{48}, 123517 (2007).
\bibitem{HoffdaSilva:2009is} 
J.M.Hoff da Silva, R.da Rocha,
``From Dirac Action to ELKO Action'',
\textit{Int.J.Mod.Phys.A}\textbf{24}, 3227 (2009).
\bibitem{Cavalcanti:2014uta}
R.T.Cavalcanti,J.M.Hoff da Silva,R.da Rocha,``VSR symmetries
in the DKP algebra: the interplay between Dirac 
and Elko spinor fields'',
\textit{Eur.Phys.J.Plus}\textbf{129}, 246 (2014).
\bibitem{Bernardini:2005wh} 
A.E.Bernardini, S.D.Leo,
``Flavor and chiral oscillations with Dirac wave packets'',
\textit{Phys.Rev.D}\textbf{71}, 076008 (2005).
\bibitem{Bernardini:2012sc}
A.E.Bernardini, R.da Rocha,
``Dynamical dispersion\\ relation for ELKO dark spinor fields'',\\
\textit{Phys.Lett.B}\textbf{717}, 238 (2012).
\bibitem{daRocha:2011yr}
R.da Rocha, A.E.Bernardini, J.M.Hoff da Silva, ``Exotic Dark Spinor Fields'', 
\textit{JHEP} \textbf{1104}, 110 (2011).
\bibitem{Rodrigues:2005yz}
W.A.Rodrigues, R.da Rocha, J.Vaz, ``Hidden\\ consequence of active local Lorentz invariance'',\\ \textit{Int.J.Geom.Meth.Mod.Phys.}\textbf{2}, 305 (2005).
\bibitem{Fabbri:2016msm}
L.Fabbri,
``A generally-relativistic gauge\\ classification of the Dirac fields'',\\ \textit{Int.J.Geom.Meth.Mod.Phys.}\textbf{13},1650078(2016).
\bibitem{Fabbri:2016laz}
L.Fabbri,
``Torsion Gravity for Dirac Fields'',\\ \textit{Int.J.Geom.Meth.Mod.Phys.}\textbf{14},1750037(2017).
\bibitem{Fabbri:2017pwp}
L.Fabbri, ``General Dynamics of Spinors'',\\ 
\textit{Adv. Appl. Clifford Algebras}\textbf{27}, 2901 (2017).
\bibitem{Fabbri:2018crr}
L.Fabbri, ``Covariant inertial forces for spinors'',\\ 
\textit{Eur.J.Phys.C}\textbf{78}, 783 (2018).
\bibitem{Fabbri:2019kfr}
L.Fabbri, ``Polar solutions with tensorial connection of\\ the spinor equation'', \textit{Eur.Phys.J.C}\textbf{79}, 188 (2019).
\end{thebibliography}
\end{document}